# Methods for Increasing the Resistance of Cryptographic Designs against Horizontal DPA Attacks

Ievgen Kabin, Zoya Dyka, Dan Kreiser and Peter Langendoerfer

IHP, Im Technologiepark 25, 15236, Frankfurt (Oder), Germany
{kabin, dyka, kreiser, langendoerfer}@ihp-microelectronics.com

**Abstract.** Side channel analysis attacks, especially horizontal DPA and DEMA attacks, are significant threats for cryptographic designs. In this paper we investigate to which extend different multiplication formulae and randomization of the field multiplier increase the resistance of an ECC design against horizontal attacks. We implemented a randomized sequence of the calculation of partial products for the field multiplication in order to increase the security features of the field multiplier. Additionally, we use the partial polynomial multiplier itself as a kind of countermeasure against DPA attacks. We demonstrate that the implemented classical multiplication formula can increase the inherent resistance of the whole ECC design. We also investigate the impact of the combination of these two approaches. For the evaluation we synthesized all these designs for a 250 nm gate library technologies, and analysed the simulated power traces. All investigated protection means help to decrease the success rate of attacks significantly: the correctness of the revealed key was decreased from 99% to 69%.

## 1    Introduction

Wireless Sensor Networks (WSNs) and the Internet of Things (IoT) are emerging technologies and are used in application fields such as telemedicine, automation control and monitoring of critical infrastructures. These application fields require the data to be kept confidential and/or to ensure the integrity of transmitted data.

   RSA and Elliptic curve cryptography (ECC) are asymmetric cryptographic approaches. Both can be applied not only for encryption and decryption of messages but also for digital signature operations and for key exchange. To reduce the time and energy consumption of computation, asymmetric cryptographic algorithms are implemented in hardware, as cryptographic accelerators. The area of cryptographic accelerators defines its production costs as well as its energy consumption per clock cycle, so it has to be as small as possible. As ECC uses by far smaller keys than RSA, it provides an energy efficient kind of public key cryptography and is well suited for WSNs and for the IoT. In this type of networks the risk of side channel analysis (SCA) attacks needs to be taken serious. Due to the need of saving energy i.e. sleep-ing intervals and due to the nature of wireless connections devices can be stolen unnoticed, analysed in a labor and brought back. So, the devices or better the implementations of cryptographic operations need to be as resistant to SCA attacks as possible.

In ECC each cryptographic key pair consists of a private and a public component. As the security of the ECC is based on keeping the private key secret the goal of an attacker is to reveal this key. The most often applied attacks are power analysis (PA) attacks or electromagnetic analysis (EMA) attacks. The attacker measures the current through the crypto-accelerator or its electromagnetic emanation while a cryptographic operation using the private key or other sensitive data is performed. For ECC the core operation is the elliptic curve point multiplication with a scalar, denoted as *kP* operation. *P* is a point of the elliptic curve (EC) and *k* is a scalar. For the ECDSA signature generation [1] the critical operation is a *kG* multiplication, i.e. a multiplication of the EC basis point *G* with a random number *k*. If an attacker can reveal the scalar *k*, the private key *Key* used for a signature generation can be easy calculated as follows:

$$Key = \frac{s \cdot k - e}{r} \mod \varepsilon$$

Here *e* is a hash value of the message to be signed; numbers *r* and *s* are components of the digital signature and *ε* is the order of the EC basis point *G*, respectively to [1]-[2]. The numbers *r, s* and the message itself are transmitted to a receiver, i.e. the attacker knows these numbers. Additionally, the point *G* and its order *ε* are parameters of the EC, i.e. they are known to the attacker, as well.

*kP* algorithms implemented in hardware process the scalar *k* bitwise. Thus, the processing of each key bit takes a certain time, here denoted as a slot. The shape of a slot in a measured trace depends on the circuit of the ECC design, on the value of the processed key bit and on the data processed in the slot. This means that the measured traces can be used for revealing the scalar *k*. Horizontal attacks [3], i.e. attacks based on a statistical analysis of a single trace, are significant threats for cryptographic devices, especially due to the fact that the traditional randomization methods such as randomization of the scalar *k*, blinding of the EC point *P* or randomization of the projective coordinates of point *P* [4] do not provide any kind of protection.

### 1.1 Contribution of This Paper

In this paper we report on the impact of using a classical multiplication formula on the success of the low-cost horizontal DPA attack, that is described in [5]-[8]. Additionally we randomized the sequence of calculation of partial products by each field multiplication as described in [5]. The idea to randomize the sequence was proposed in [15]-[16]. In [15]-[16] it was evaluated only for a field multiplier. As reported in [5] this randomization increases the resistance of ECC designs against horizontal attacks but is not sufficient as a single countermeasure. In consequence we combined this approach with applying of the classical multiplication formula for the calculation of partial products. To evaluate these countermeasures we performed the horizontal attack as described in [5] against different ECC designs using simulated power traces for 250 nm gate library technologies.

The rest of this paper is structured as follows. In section 2 we describe investigated *kP* designs. In section 3 we explain shortly how we performed the horizontal low-cost DPA attack using the difference of the means and how we evaluated the success of attacks. The results of the attacks are evaluated in section 4. Conclusions are given in section 5.

## 2    Investigated *kP* Designs

1987 Montgomery proposed an algorithm for the *kP* calculation [9]. 1998 Lopez and Dahab showed that the Montgomery *kP* algorithm can be performed using only the *x*-coordinate of the point *P* if *P* is a point on EC over *GF($2^n$)* [10]. Additionally, they proposed to use special projective coordinates of the EC point *P* to avoid the most complex operation, i.e. the division of elements of Galois fields. These optimizations reduced the execution time and the energy consumption of the *kP* calculation significantly. The Montgomery algorithm using projective Lopez-Dahab coordinates is a time and energy efficient solution and due to this fact this algorithm is the one mostly used for implementing the EC point multiplication in hardware. The most referenced version of the Montgomery *kP* algorithm is [11]. The *kP* operation according to this algorithm, can be performed using only 6 multiplications, 5 squarings and 3 additions of Galois field elements for each key bit, except of the most significant bit $k_{l-1}=1$. The length of the operands depends on the chosen security level. We experimented with a *kP* design for EC *B-233*, recommended by NIST [1]. The maximal length *l* of operands is up to 233 bits in our designs.

The Montgomery *kP* algorithm has the same sequence of operations for the processing of each key bit, independently of its value. Such implementations are resistant against SPA attacks. A possibility to increase the inherent resistance of the Montgomery *kP* implementations against SCA attacks (not only against simple ones), is to increase the noise level in the analysed power profile. As reported in [7] the field multiplier can be the source of the noise if itself is resistant against SCA attacks. The *write to register* operations are most analysed ones while an attack is performed. If these operations are executed in parallel to the field multiplications the analysis becomes by far more complex. Thus, implementations exploiting parallel execution of operations of the *kP* algorithm are inherently more resistant against PA attacks. Additionally, the execution of many operations in parallel reduces the execution time of the cryptographic operations and increases the efficiency of the design.

Our *kP* design is a balanced and efficient implementation of the *kP* algorithm based on Algorithm 2 published in [7], that is a modification of the Montgomery *kP* algorithm.

### 2.1    Basic Design: Balanced, Efficient, Resistant against SPA and HCCA

The structure of our *kP* designs is shown in Fig. 1.

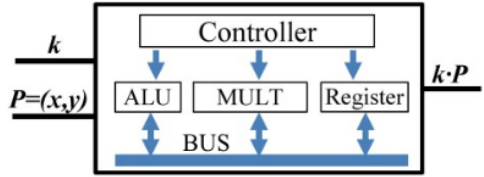

**Fig. 1.** Structure of our *kP* design.

The block Controller manages the sequence of the field operations. It controls the data flow between the other blocks and defines which operation has to be performed in the current clock cycle. Depending on the signals of the Controller the block ALU performs addition or squaring of its operands. Our design comprises of only one block MULT to calculate the product of 233 bit long operands. The multiplication is the most complex field operation. In our implementation it takes 9 clock cycles to calculate the product according to a fixed calculation plan using the iterative Karatsuba multiplication method as described in [12]. In each of the 9 clock cycles one partial polynomial product of two 59 bit long operands $A_j$ and $B_j$ (with $1 \leq j \leq 9$) is calculated and accumulated to the product including reduction.

Fig. 2 shows the structure of our field multiplier for 233 bit long operands. It consists of a Partial Multiplier (PM) for 59 bit long operands. The field multiplier takes 9 clock cycles to calculate the product using a 59 bit partial multiplier. The PM takes 1 clock cycle to calculate the polynomial product of 59 bit long operands and is implemented as a combination of 3 multiplication methods (MMs). The 2-segment iterative Karatsuba multiplication formula [14] was applied for 60-bit long multiplicands. The gate complexity of this multiplier is $GC_{2m}=3 \cdot GC_m+(7m-3)_{XOR}$. Here $m$ is the length of segments $m=60/2=30$ and $GC_m$ is the gate complexity of the internal $m$-bit partial multipliers. Thus, the 59 bit partial multiplier contains of 3 internal multipliers: two of them for 30 bit long operands and one multiplier for 29 bit long operands.

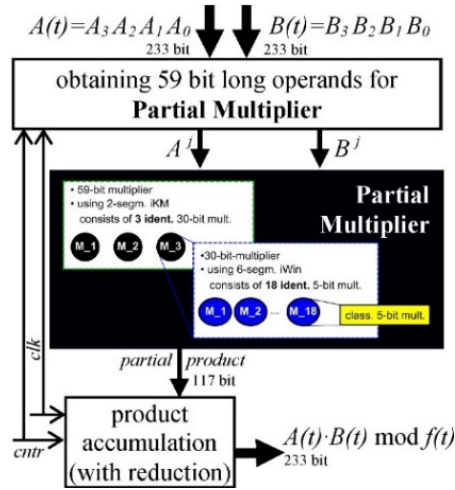

**Fig. 2.** Structure of the field multiplier for 233 bit long operands.

All these internal multipliers are implemented identically, using the 6-segment iterative Winograd multiplication formula [14], which gate complexity is: $GC_{6m}=18 \cdot GC_m+(72m-19)_{XOR}$, with $m=30/6=5$ bits. Corresponding to the 6-segment iterative Winograd multiplication formula the 30-bit multiplier consists of 18 internal multipliers of 5-bit long operands. Each of these small multipliers was implemented using the classical multiplication formula with $n=5$:

$$C = A \cdot B = \sum_{i=0}^{2n-2} c_i \cdot t^i, \text{ with } c_i = \bigoplus_{i=k+l} a_k \cdot b_l, \forall k,l < n \ . \quad (1)$$

The gate complexity of each of the 5-bit classical multiplier is $GC_5=25_\&+16_{XOR}$, i.e. $5^2=25$ AND gates and $4^2=16$ XOR gates.

Applying the described combination of 3 multiplication methods, the gate complexity of the 59-bit partial multiplier is $GC_{59}=1350_\&+2094_{XOR}$. This PM is area optimized for the applied technology. The minimal possible processing time per key bit in our implementation is equal to the time needed to execute 6 field multiplications (here taking 9 clock cycles each), i.e. 54 clock cycles. This is achieved by implementing all arithmetic and *write to register* operations in parallel to the multiplications.

Our Basic Design is inherent resistant against Horizontal Collision Correlation Analysis (HCCA) attacks introduced in [13]. This is achieved by using the iterative 4-segment Karatsuba MM:

— the number of calculated partial products is small (only 9);
— the length of operands for each partial multiplication is big, $w=59$;
— operands for partial product calculation using the iterative Karatsuba MM are different, that is not the case if the classical MM as assumed in [13] is applied.

### 2.2 Design with Randomized Sequence of PMs

Usually a field product of long operands is calculated as a sum of partial products of partial operands of smaller length. In each clock cycle one partial product is calculated and accumulated according to a fixed scheduled calculation sequence often denoted as accumulation plan. In [15]-[16] it was proposed to randomize the calculation sequence of the partial products, i.e. to re-schedule the calculation plan for each new field multiplication with the goal to increase the resistance of the field multiplier against SCA attacks. In [5] this method was evaluated as a countermeasure against horizontal DPA attacks using an execution of a *kP* operation. If the multiplication formula consists of *n* partial products, there exist *n!* different permutations of this sequence. One of these permutations can be selected randomly for the calculation of each field multiplication.

The field reduction has to be applied to the accumulation register and can be performed either once per field multiplication or after calculating of each partial product. The latter design consumes more power for the calculation of the filed product but the power shape of such a multiplication is more random. The partial reduction after each calculation of a partial product was implemented not only in [15]-[16] but also in the design reported in [17]. All designs investigated here perform the reduction of the product after the calculation of each partial product to increase the noise and to reduce the success of SCA attacks.

In our Basic Design the multiplication formula contains 9 partial products of 59 bit long operands, i.e. for each new field product calculation one out of 9! possible calculation sequences can be selected randomly.

Table 1 gives an overview of implementation details of our design and the implementation described in [15]-[16].

**Table 1.** Overview of implementation details of two randomized multipliers

| parameters | design [15]-[16] | our design |
|---|---|---|
| field multiplier for | $GF(2^r)$-elements | $GF(2^r)$-elements |
| lengths of multiplicands | $r=192$ | $r=233$ |
| irreducible polynomial | not given | $f(t)=t^{233}+t^{74}+1$ |
| #segments | 6 | 4 |
| Applied multiplication formula | $eMSK_{6=2*3}$ | 4-segment iterative Karatsuba MM |
| Partial multiplier for | 32 bit long operands | 59 bit long operands |
| #partial multiplications | 18 | 9 |
| #possible permutations | 18! | 9! |

The smaller number of possible permutations in our design compared to the one described in [15]-[16] means that our multiplier is more vulnerable to collision-based attacks. They are a kind of vertical attacks and can be prevented using traditional randomization countermeasures [4]. In this work we concentrate on the prevention of horizontal DPA attacks. The area and energy consumption of a partial multiplier for 59-bit long operands are significantly higher than those of a multiplier for 32 bit long operands. Thus, the 59-bit partial multiplier can be more effective as a noise source and by that as a means against horizontal DPA attacks.

### 2.3 Design with Classical PM

The next design we used in our experiments was Basic Design (see section 2.1) but here the partial multiplier was implemented using the classical multiplication formula only, i.e. it implements formula (1) for the length of the partial multiplicands $n=59$.

The gate complexity (GC) of this multiplier, i.e. the amount of *AND* and *XOR* gates which are necessary to implement its functionality corresponding to formula (1) is $n^2$ *AND* gates and $(n-1)^2$ *XOR* gates, i.e.: $GC_{59}=3481_{\&}+3364_{XOR}$.

The gate complexity of such a multiplier is the biggest one of all potential multipliers. All other multiplication methods, like Karatsuba or eMSK multiplication formulae, were developed with the goal to reduce the (gate) complexity of the classical multiplication formula. On the one hand the gate complexity is a disadvantage of the classical multiplication method because it results in an increased chip area, price and energy consumption. But on the other hand the increased energy consumption and especially its fluctuation mean an increased noise level for an attacker, if it analyses the activity of the other blocks. Due to this fact, using the classical MM for the implementation of the partial multiplier can be an advantage, because it increases the inherent robustness of *kP* designs against SCA attacks.

### 2.4  Design with Classical PM and Randomized Sequence of PMs

The PM was implemented using the classical MM as it was done for the design introduced in section 2.3. The sequence of the partial multiplications was re-scheduled for each new field multiplication as described for the design introduced in section 2.2. Thus, our 4$^{th}$ implemented design is a combination of both approaches for increasing the resistance against SCA attacks.

## 3  Horizontal DPA Using the Difference of the Means

To perform a horizontal DPA attack we prepared the power traces as follows:

– We fragmented the power trace in time slots. Each time slot corresponds to the processing of a bit of the used scalar *k*. In the rest of the paper we denoted the scalar *k* as key. For our analysis we selected only time slots where key bits were processed in the main loop of the Montgomery *kP* algorithm. In our experiments with a 232 bit long randomly generated key *k* the slots correspond to $k_j$ with $0 \leq j \leq 229$. I.e. we excluded the processing of the two most significant bits of the key from the analysis. In our implementation one time slot consists of 54 clock cycles.
– We averaged the power per clock cycle to represent the clock cycle in the analysis by only one power value, i.e. we compressed the trace.

We performed our horizontal DPA attack using the difference of the means applied to the compressed traces as follows:

1. Using the 230 time slots we calculated the arithmetical mean of all values with the same number *i*, which is the number of the clock cycle *(1≤i≤54)* within the time slot, and different number *j*:

$$\overline{p^i} = \frac{1}{230} \sum_{j=0}^{229} p^i_j \qquad (2)$$

Thus, the 54 calculated values $\overline{p^i}$ define the mean power profile of slots.

2. For each *i* we obtained one key candidate $k^{candidate\_i}$ using the following assumption: the *j*$^{th}$ bit of the key candidate is 1 if in the slot with number *j* the value with number *i* – i.e. the value $p^i_j$ – is smaller than or equal to the average value $\overline{p^i}$. Else the *j*$^{th}$ bit of the *i*$^{th}$ key candidate is 0:

$$k^{candidate\_i}{}_j = \begin{cases} 1, & if \quad p^i_j \leq \overline{p^i} \\ 0, & if \quad p^i_j > \overline{p^i} \end{cases} \qquad (3)$$

To evaluate the success of the attack we compared all extracted key candidates with the scalar *k* that was really processed. For each key candidate we calculated its relative correctness as follows:

$$\delta_1 = \frac{number\_of\_correct\_extracted\_bits\_of\_k^{candidate\_i}}{230} \cdot 100\% \qquad (4)$$

The range of the correctness $\delta_1$ is between 0 and 100%.

If a key candidate was extracted with a correctness close to 0 percent, it means that our assumption in equation (3) is wrong and the opposite assumption will be correct. Thus, the relative correctness $\delta_1 = 0$ of the key candidate obtained using assumption (3) will correspond to correctness $\delta_1 = 100$ percent if the opposite assumption is used. Taking this fact into account we can calculate the correctness as follow:

$$\delta = 50\% + |50\% - \delta_1| \qquad (5)$$

Thus, we define the correctness as a value between 50% and 100%. For the attacker the worst-case of the attack results is a correctness of 50 percent which means the *difference of means* test cannot even provide a slight hint whether the key bit processed is more likely a '1' or a '0'. The worst-case from the attacker's point of view is the ideal case from the designer's point of view. We denote it as the "ideal case" in the rest of the paper.

Fig. 3 shows attack results i.e. relative correctness $\delta$ for the key candidates extracted using PTs of our Basic Design, simulated for the 250 nm technology. In order to demonstrate that well-known countermeasures [4] are not effective against horizontal DPA, we applied point blinding, key randomization and a combination of both as countermeasures with the goal to randomize the data processed in our Basic Design. The red bars show the result of the attack for the Basic Design without randomized inputs. The green, the yellow and the black bars show the analysis results when traditional countermeasures [4] are applied.

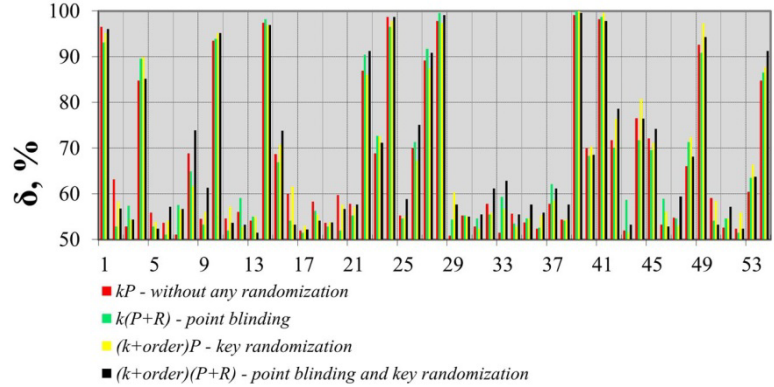

**Fig. 3.** Results of attacks against *kP* executions with and without traditional randomization countermeasure. All PTs were simulated for our Basic Design using the 250 nm technology.

Fig. 4 shows all key candidates given in Fig. 3 sorted in descending order of correctness. According to that each key candidate got a new index displayed at the *x* axis. This representation helps to compare the analysis results.

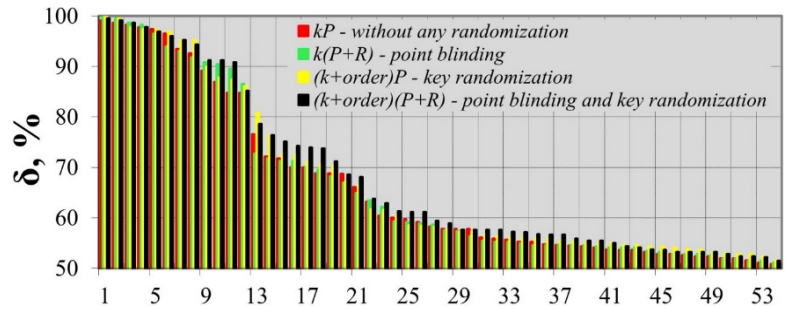

**Fig. 4.** Correctness of key candidates from Fig. 3 sorted in descending order; according to that each key candidate got a new index displayed at the *x* axis. The one with the highest correctness is now number 1, the one with the lowest number 54. The blue horizontal line at 50 percent shows the ideal case.

Comparing attack results displayed in Fig. 3 and Fig. 4 shows clearly that the traditional randomization countermeasures do not provide any protection against horizontal SCA attacks.

## 4 Discussion of the Results for Investigated Designs

In this section we discuss the analysis results of the power shape randomization strategies introduced in section 2. We synthesized the 4 designs described in section 2 for a 250 nm technology. Then we simulated the designs using PrimeTime [18] to get power traces which we then analysed to evaluate the effectiveness of the randomiza-tion strategies for the complete *kP* designs. All 4 power traces were simulated using the same inputs, i.e. the key *k* and EC point *P*. Fig. 5 shows the analysis results.

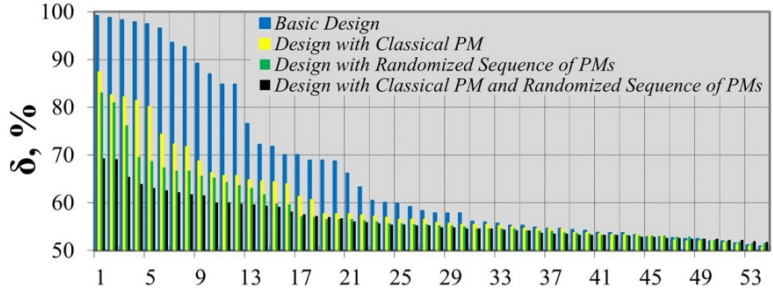

**Fig. 5.** Attack results: correctness of the extracted keys sorted in descending order.

The results of the analysis show that the implementation of the partial multiplier using the classical multiplication formula has a significant impact on the resistance of the *kP* design against horizontal DPA (see yellow bars in Fig.5). This effect is similar to the randomization of the calculation sequence of partial products if the PM is implemented as an area-optimized combination of MMs (see green bars). Both strategies combined, i.e. applying a randomized sequence of PMs and implementation of PM using the classical MM increases this effect significantly: the correctness of the extraction was decreased from 99% for our Basic Design (see blue bars) to 69% (see black bars) that is a significant improvement of the design's resistance against the applied horizontal DPA attack.

## 5  Conclusion

In this paper we showed that traditional countermeasures such as point blinding and key randomization provide almost no protection against horizontal DPA attacks (see Fig. 3 and Fig. 4 in section 3). In order to prevent horizontal DPA attacks from being successful we investigated alternative means to increase the resistance of the *kP* designs: randomizing the calculation sequence of the partial products and implementing the partial multiplier using the classical multiplication formula. We showed that the impact of both countermeasures on the success of horizontal DPA attacks is similar. Especially a combination of these approaches can significantly increase the inherent resistance of ECC designs against horizontal attacks: the correctness of the revealed key was decreased from 99% to 69% (see Fig. 5 in section 4).

**Acknowledgments**. The work presented here was partly supported by the German Ministry of Research and Education (BMBF) within the ParSec project, grant agreement no. 16KIS0219K.

# References

1. Federal Information Processing Standard (FIPS) 186-4, Digital Signature Standard; Request for Comments on the NIST-Recommended Elliptic Curves: 2015.
2. Johnson, D., Menezes, A., Vanstone, S.: The Elliptic Curve Digital Signature Algorithm (ECDSA). IJIS. 1, 36–63 (2001).
3. Clavier, C., Feix, B., Gagnerot, G., Roussellet, M., Verneuil, V.: Horizontal correlation analysis on exponentiation. In: In ICICS 2010, volume 6476 of LNCS. pp. 46–61. (2010).
4. Coron, J.-S.: Resistance Against Differential Power Analysis For Elliptic Curve Cryptosystems. In Proceedings of CHES'99. pp. 292–302.
5. Kabin, I., Dyka, Z., Kreiser, D., Langendoerfer, P.: Evaluation of Resistance of ECC Designs protected by Different Randomization Countermeasures against Horizontal DPA Attacks. In Proceedings of IEEE East-West Design Test Symposium (EWDTS2017).
6. Kabin, I., Dyka, Z., Kreiser, D., Langendoerfer, P.: Attack against Montgomery kP Implementation: Horizontal Address-Bit DPA?. In Proceedings of the WiP Session of Euromicro Conference on Digital System Design (DSD2017).
7. Dyka, Z., Bock, E.A., Kabin, I., Langendoerfer, P.: Inherent Resistance of Efficient ECC Designs against SCA Attacks. In: 2016 8th IFIP International Conference on New Technologies, Mobility and Security (NTMS). pp. 1–5 (2016).
8. Kabin, I., Dyka, Z., Kreiser, D., Langendoerfer, P.: On the Influence of Hardware Technologies on the Vulnerability of Protected ECC Implementations. In Proceedings of the WiP Session of Euromicro Conference on Digital System Design (DSD2016).
9. Montgomery, P.L.: Speeding the Pollard and elliptic curve methods of factorization. Math. Comp. 48, 243–264 (1987).
10. López, J., Dahab, R.: Fast Multiplication on Elliptic Curves Over GF(2m) without precomputation. In Proceedings of CHES'99. pp. 316–327.
11. Hankerson, D., Hernandez, J.L., Menezes, A.: Software Implementation of Elliptic Curve Cryptography over Binary Fields. In: Cryptographic Hardware and Embedded Systems — CHES 2000. pp. 1–24. Springer, Berlin, Heidelberg (2000).
12. Dyka, Z., Langendoerfer, P.: Area efficient hardware implementation of elliptic curve cryptography by iteratively applying Karatsuba's method. In: Design, Automation and Test in Europe. p. 70–75 Vol. 3 (2005).
13. Bauer, A., Jaulmes, E., Prouff, E., Wild, J.: Horizontal Collision Correlation Attack on Elliptic Curves. In: Selected Areas in Cryptography -- SAC 2013. pp. 553–570 (2013).
14. Dyka, Z. Analysis and prediction of area- and energy-consumption of optimized polynomial multipliers in hardware for arbitrary GF($2^n$) for elliptic curve cryptography. Dissertation thesis, BTU Cottbus-Senftenberg (2013). https://opus4.kobv.de/opus4-btu/frontdoor/index/index/docId/2634.
15. Madlener, F., Sötttinger, M., Huss, S.A.: Novel hardening techniques against differential power analysis for multiplication in GF(2n). In: 2009 International Conference on Field-Programmable Technology. pp. 328–334. IEEE (2009).
16. Stöttinger, M., Madlener, F., Huss, S.A.: Procedures for Securing ECC Implementations Against Differential Power Analysis Using Reconfigurable Architectures. In: Platzner, M., Teich, J., and Wehn, N. (eds.) Dynamically Reconfigurable Systems. pp. 395–415 (2010).
17. Dyka, Z., Wittke, C., Langendoerfer, P.: Clockwise Randomization of the Observable Behaviour of Crypto ASICs to Counter Side Channel Attacks. In: 2015 Euromicro Conference on Digital System Design. pp. 551–554 (2015).
18. Synopsis. PrimeTime http://www.synopsys.com/Tools/